\documentclass[conference]{IEEEtran}
\IEEEoverridecommandlockouts
\usepackage{cite}
\usepackage{amsmath,amssymb,amsfonts}
\usepackage{algorithm}
\usepackage{algorithmic}
\usepackage{graphicx}
\usepackage{textcomp}
\usepackage{xcolor}
\usepackage{balance}

\begin{document}

\title{A~GNN-Enhanced Reinforcement Learning Framework for Emergency Communications in ORAN-based Non-Terrestrial Networks}

\author{\IEEEauthorblockN{Md. Thouhidur Rahman\IEEEauthorrefmark{1}\IEEEauthorrefmark{3}, Mustapha Benjillali\IEEEauthorrefmark{1}\IEEEauthorrefmark{2}, Halim Yanikomeroglu\IEEEauthorrefmark{3}, and~Samir~Saoudi\IEEEauthorrefmark{1}}

\IEEEauthorblockA{\IEEEauthorrefmark{1} IMT-Atlantique, Lab-STICC, UMR CNRS 6285, Brest, France.}

\IEEEauthorblockA{\IEEEauthorrefmark{2} Communications Systems Department, INPT, Rabat 10100, Morocco.}

\IEEEauthorblockA{\IEEEauthorrefmark{3} Carleton-NTN Lab, Department of Systems and Computer  
Engineering, Carleton University, Canada.}

\IEEEauthorblockA{Emails: \{md-thouhidur.rahman,samir.saoudi\}@imt-atlantique.fr, benjillali@ieee.org, halim@sce.carleton.ca}

\thanks{This work is funded in part by the European Union’s Horizon Europe Research and Innovation Program under Grant Agreement No. 101126644.}
}

\maketitle

\begin{abstract}
During disaster scenarios and periods of extreme data demand in next-generation wireless communications, conventional terrestrial networks (TNs) often become unreliable or fail entirely, leading to critical service disruptions. In such contexts, non-terrestrial networks (NTNs) emerge as a promising solution to ubiquitous and resilient connectivity. Furthermore, the open radio access network (ORAN) paradigm facilitates network disaggregation and flexible functional splitting among its key components, namely the central unit (CU), distributed unit (DU), and radio unit (RU), which can be deployed across heterogeneous NTN platforms according to service requirements. However, this flexibility introduces significant challenges in terms of network complexity and real-time control. To address these challenges, this paper proposes an intelligent ORAN-enabled NTN framework for emergency communication scenarios. The proposed system leverages graph neural networks (GNNs) to model the dynamic network topology and employs a reinforcement learning (RL)-based Q-learning algorithm, formulated as a Markov decision process (MDP), to enable adaptive and real-time network control. In this framework, network nodes are treated as states, and optimal decisions are learned based on system dynamics. The spatial distribution of user equipment (UE) is modeled using an inhomogeneous Poisson point process (IPPP) with a rejection sampling technique, capturing realistic user density variations. Simulation results demonstrate that the proposed GNN-enhanced RL approach significantly improves network performance in terms of latency and service reliability, thereby enabling efficient and robust operation under emergency conditions.
\end{abstract}

\begin{IEEEkeywords}
Emergency Communication, GNN, Latency, NTN, ORAN, RL.
\end{IEEEkeywords}

\section{Introduction}
Recent studies indicate that conventional terrestrial base stations (TBSs) often fail or struggle to provide reliable services during natural disasters, in sparsely populated rural areas, and under high data rate demands such as virtual reality, ultra-high-definition video streaming, autonomous driving, remote surgery, cloud gaming, smart city applications, and military operations \cite{rahman2025fanet}, \cite{karaman2025demand} and \cite{karaman2026haps}. In this context, non-terrestrial networks (NTNs), including unmanned aerial vehicles (UAVs), high-altitude platform stations (HAPS), and satellites, have emerged as a promising solution to provide flexible and resilient connectivity \cite{nguyen2024emerging}, and \cite{ince2026aoi}. Meanwhile, the open radio access network (ORAN) paradigm enables network disaggregation and functional splitting through its key components—the central unit (CU), distributed unit (DU), and radio unit (RU)—which can be dynamically deployed across heterogeneous NTN platforms \cite{mahboob2025transforming,rihan2023ran,rahman2026performance}. Nevertheless, such distributed architectures introduce significant complexity in network management and resource optimization \cite{deng2026ai}. To address these challenges, this paper proposes an intelligent ORAN framework for NTNs, leveraging graph neural networks (GNNs) to model the network topology and reinforcement learning (RL) to enable adaptive, real-time control based on user equipment (UE) distribution and service requirements. The proposed approach aims to optimize network performance in terms of latency,  particularly under dynamic and emergency conditions. NTNs can dynamically update their positions, while ORAN can adapt its component resources according to varying service requirements.

\begin{figure*}[!t]
\centering
\includegraphics[width=6in]{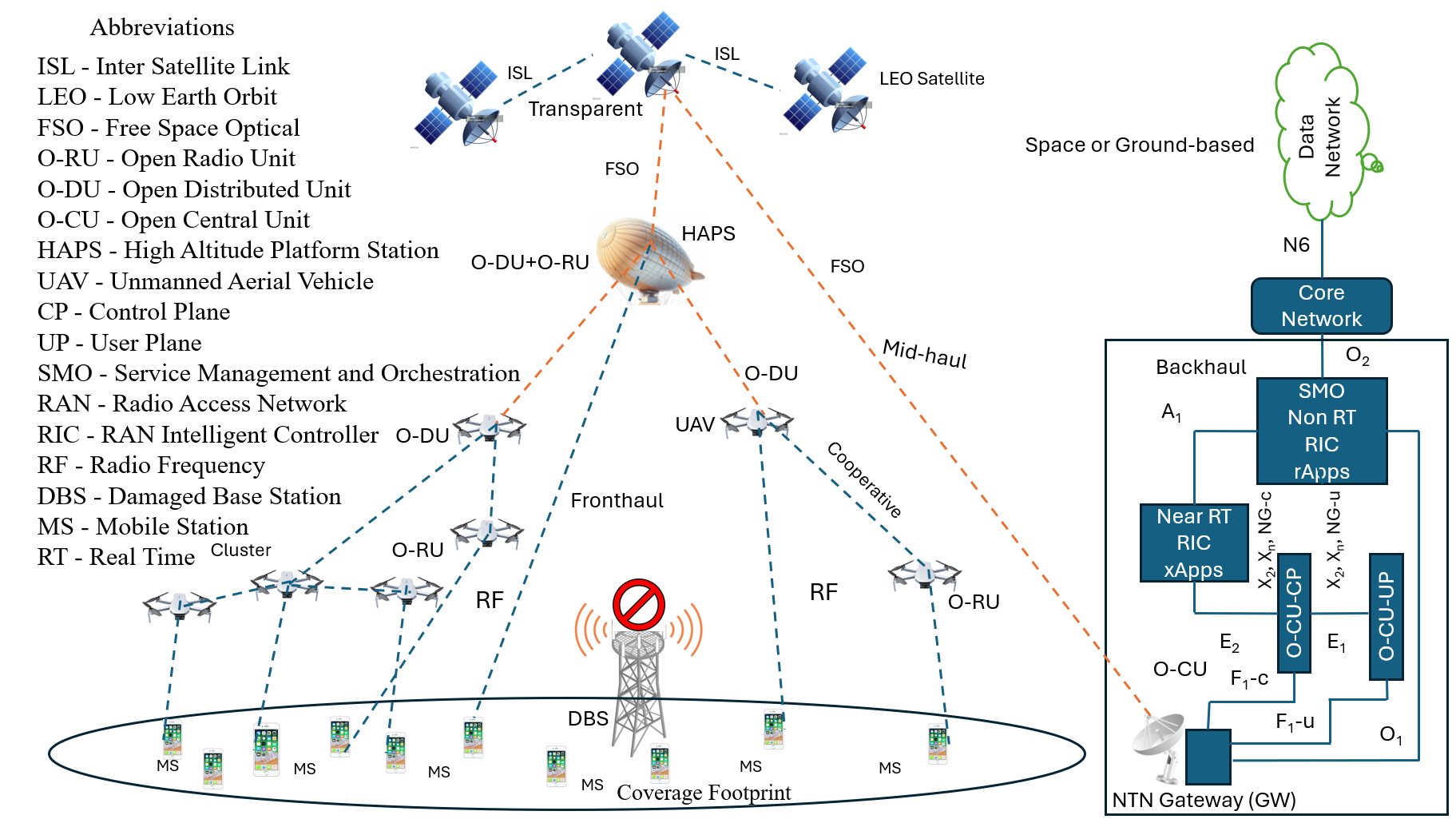}
\caption{Schematic NTNs architecture for emergency communication.}
\label{fig_1}
\end{figure*}

The schematic NTN architecture for emergency communications is illustrated in Fig. 1. The ORAN components are distributed across different platforms according to the service requirements. The CU is deployed at the NTN gateway (GW) and connected to both the core network and the data network. Recently, space-based data centers have gained significant attention \cite{periola2020space}; therefore, the CU-GW can be implemented either in space or on the ground \cite{lan2026architecture}. However, from a network perspective, space-based deployment is more suitable for supporting emergency scenarios. Low Earth orbit (LEO) satellites are connected via backhaul links using free-space optical (FSO) communication, enabling them to operate as transparent platforms. These LEO satellites are interconnected through inter-satellite links (ISLs), allowing seamless communication across the network. Furthermore, LEO satellites can connect to HAPS via FSO links. In particular, LEO satellites can sense the locations of UEs associated with TBSs through connectivity with the TBSs \cite{kanani2025optimizing}. When a failure or service degradation is detected, NTN resources are dynamically deployed. Initially, HAPS platforms carrying open DU (O-DU) and open RU (O-RU) functionalities are activated. Due to their quasi-stationary nature, HAPS can efficiently maintain FSO links with LEO satellites. Subsequently, additional O-RU or O-DU units can be deployed based on user density and demand. Finally, all physical network components are monitored through a GNN digital twin framework, which enables intelligent network control using GNN-enhanced RL algorithms. 

Due to the flexibility and advantages of ORAN–enhanced NTNs controlled autonomously, a significant number of studies have been conducted. A GNN-enhanced multi-agent deep reinforcement learning (MADRL) approach for dynamic packet routing \cite{ran2024fully} and a latency-tolerant algorithm for heterogeneous NTNs were analyzed in~\cite{deng2026distributed}. In those works, dynamic routing in NTNs was formulated as a partially observable Markov decision process (MDP), and residual shortest path hops were used for routing selection. The integration of terrestrial networks (TNs) and NTNs was investigated in \cite{tsegaye2024graph}, where the open CU (O-CU) was deployed in the TN, while the O-DU was distributed across NTN platforms. A GNN-based graph convolutional network (GCN) was utilized for fault detection and efficient routing path selection to mitigate failures. ORAN NTN functional splitting optimization using a deep Q-learning network (DQN)-based RL, adapted to varying traffic conditions, was studied in \cite{shahabi2024energy}. An actor–critic (A2C) algorithm combined with a sequence-to-sequence DRL model for optimizing energy consumption in ORAN functional splitting was proposed in \cite{amiri2023energy}, and \cite{pamuklu2021reinforcement}. Dynamic functional splitting using RL to optimize energy consumption with renewable energy sources was presented in~\cite{pamuklu2021reinforcement}. The distribution of ORAN components across multiple UAVs for 5G network task offloading using a DQN-based RL algorithm was investigated in \cite{pham2022ran}. Recently, the architectural design and resource allocation for ORAN in space were developed in \cite{lan2026architecture}. ORAN mobility management based on a GNN and a link prediction method was investigated in \cite{bermudez2025graph}.

In view of the limitations and challenges identified in the existing literature on NTN-ORAN systems, the main contributions of this work are summarized as follows:
\begin{itemize}
    \item To address the major challenges of ORAN-based NTNs, including dynamic topology management and intelligent network control, a GNN-enhanced RL framework is proposed for ORAN functional splitting in emergency communication scenarios.
    \item Based on the damaged base station (DBS) locations, the UE distribution is modeled using an IPPP with a rejection sampling technique, and an efficient, scalable end-to-end NTN architecture is developed to support emergency communications.
    \item A GNN-based network model represents NTN ORAN nodes and communication links as a graph to learn real-time network topology and link characteristics. Based on the learned graph representation, a Q-learning-based RL mechanism within an MDP framework optimizes functional split selection and resource management through an appropriate reward function while satisfying the target latency.
    \item Simulation results demonstrate that the proposed framework effectively learns optimal control policies, achieving the target latency with up to 90\% successful action rates and ensuring reliable and efficient network performance.
\end{itemize}

The remainder of this article is organized as follows. Section~II discusses the GNN framework for the NTN architecture. Section~III presents the RL model for network adaptation. Section~IV provides the performance analysis of the GNN-enhanced RL training. Finally, Section~V concludes the paper and outlines future research directions.

\section{GNN Framework for NTN Architecture}

Based on the physical network illustrated in Fig. 1, a digital framework is designed using a GNN. In this framework, all NTN platforms (O-DU, O-RU), UEs, the CU GW, the core network, and the data network are modeled as nodes. The efficient and flexible Python library NetworkX is used to create and manipulate the graph structure. Communication links, such as FSO and radio frequency (RF), are represented as edges of the graph. Initially, an empty graph is constructed, and nodes are added sequentially. Each node is associated with several attributes, including the type of NTN platform, type of ORAN component, node position, neighboring nodes, neighbor types, and edge types (i.e., link types). All node features are summarized in Table I. 

Subsequently, the positions of the NTN platforms are determined based on the location of the damaged base station (DBS). The HAPS (DU+RU) is positioned directly above the DBS, while the LEO satellite (relay) is placed above the HAPS. Other NTN platforms are distributed according to the spatial density of UEs. The UE distribution is modeled using a two-dimensional Gaussian IPPP. UEs are generated via a rejection sampling method based on a Gaussian probability density function (PDF), resulting in a higher concentration near the center of the DBS area, with density gradually decreasing toward the edges. The resulting hierarchical network topology, modeled using the GNN framework, is illustrated in Fig. 2 and Fig. 3. ORAN-based NTNs involve challenges such as routing, handover, interface synchronization, link failure mitigation, and scheduling. NetworkX models the network topology by representing node and edge features, enabling the GNN to learn network characteristics such as link budgets, 3D node positions, link distances, latency, and shortest paths. The learned graph representations are then utilized by the RL agent to optimize network control policies, improving the adaptability and efficiency of ORAN-based NTNs.

\begin{table}[!t]
\caption{GNN node feature characteristics}
\label{tab:table1}
\centering
\begin{tabular}{ll}
\textbf{Feature Name} & \textbf{Attribute Characteristics} \\
\hline\hline
Network Type & O-CU/O-DU/O-RU/O-Relay \\ \hline
NTN Platforms & HAPS/UAV/LEO \\ \hline
Position & Reference to DBS 
 \\ \hline
Edge Type & FSO/RF 
  \\ \hline
Neighbor Node & HAPS/UAV/LEO/UE \\ \hline
Neighbor Node Type & UE/CU/DU/RU/Relay \\ \hline
\end{tabular}
\end{table}

To represent the user density distribution, let the density of users be \(d_{\mathrm{u}}\) according to the position of \((x_i, y_i)\). The maximum user density is \(d_{\mathrm{um}}\) if \(x_{\mathrm{o}}=0\), and \(y_{\mathrm{o}}=0\). Therefore, for a Gaussian (non-uniform) distribution, the PDF can be found~by:
\begin{equation} 
\label{deqn_ex1}
f(x_i, y_i) = \exp\left( - \left( 
\frac{(x_i - x_B)^2}{2\sigma_{x_i}^2} + 
\frac{(y_i - y_B)^2}{2\sigma_{y_i}^2}
\right) \right),
\end{equation}
with \((x_{\mathrm{B}},y_{\mathrm{B}})\) the DBS position, \(\sigma_{x_{i}}\), and \(\sigma_{y_{i}}\) are special spread constant of \(x\) coordinate and \(y\) coordinate respectively. Let the DBS coverage area be denoted by \(\mathsf{A}_{\mathrm{B}}\). Thus, the expected value of the user's positions can be illustrated as:
\begin{equation} 
\label{deqn_ex2}
\mathbb{E}[d_u(x_i, y_i)] = \iint_{\mathcal{\mathsf{A}_{\mathrm{B}}}} d_u(x_i, y_i)\, f(x_i, y_i)\, dx_i\, dy_i.
\end{equation}

In the rejection sampling method, the boundary coverage is:
\begin{equation} 
\label{deqn_ex3}
(x_i, y_i) \in [x_{\min}, x_{\max}] \times [y_{\min}, y_{\max}].
\end{equation}

If we denote the uniform distribution PDF by $f_{\text{uniform}}(x_i,y_i)$, then the accepted PDF value can be represented by: 
\begin{equation} 
\label{deqn_ex4}
f_{\text{uniform}}(x_i, y_i) < f_{\text{accepted}}(x_i, y_i),
\end{equation}
where \(f_{\text{accepted}}(x_i,y_i)\) expresses the accepted PDF of the rejection sampling method. 

\begin{table}[!t]
\caption{RL-based Model Parameters Symbolization}
\label{tab:table2}
\centering
\begin{tabular}{ll}
\textbf{Symbol} & \textbf{Representation} \\
\hline\hline
\(N_{\mathrm{HAPS}}\), \(N_{\mathrm{LEO}}\) & Number of HAPS, LEO  \\ \hline
\(N_{\mathrm{UAV}}\),\(N_{\mathrm{UE}}\) & Number of UAVs, UEs  
 \\ \hline
\(N_{\mathrm{RU}}\), \(N_{\mathrm{DU}}\) & Number of O-RU, O-DU \\ \hline
\(N_{\mathrm{NRU}}\), \(N_{\mathrm{NDU}}\) & Number of added O-RU, O-DU \\ \hline
\(N_{\mathrm{RRU}}\), \(N_{\mathrm{RDU}}\) & Number of removed O-RU, O-DU \\ \hline
\(N_{\mathrm{LN}}\), \(N_{\mathrm{RL}}\) & Number of links, relays \\ \hline
\(S_{\mathrm{t}}\) & Set of states from graph G nodes by \(s\) and \(s'\) \\ \hline
\(A_{\mathrm{c}}\) & Set of actions symbolizes by \(a\) \\ \hline
\(R(s,a,s')\) & Reward for each state \(s\), action \(a\), and next state \(s'\) \\ \hline
\(P_{\mathrm{T}}(s,a,s')\) & Transition probability for each state \\ \hline
\(\tau\) & Policy gradiant from one state to another \\ \hline
\({\tau}^*(s)\) & Optimal policy state \\ \hline
\(\gamma\), \(\alpha\) & Discount factor, Learning rate \\ \hline
\(Q(s,a)\) & \(Q\) value of state \(s\) and action \(a\) \\ \hline
\end{tabular}
\end{table}

\section{RL Model for Network Adaptation}
To implement the RL method, the elements of an MDP must be defined. All symbols and notations are summarized in Table II. The Q-learning-based dynamic learning process is presented in Algorithm 1.  At each state, the agent observes the entire graph, including the number of nodes and edges, such as \(N_{\mathrm{LEO}}\), \(N_{\mathrm{HAPS}}\), \(N_{\mathrm{UE}}\), \(N_{\mathrm{RU}}\), \(N_{\mathrm{DU}}\), \(N_{\mathrm{LN}}\), and \(N_{\mathrm{RL}}\). The action set \(A_{\mathrm{t}}\) includes operations such as adding or removing nodes (e.g., UEs, DU-UAVs, and RU-UAVs) and links, as well as updating node positions based on UE requirements and target latency constraints. The reward function is defined based on the effectiveness of the selected action. A negative reward is assigned when the average latency is high, while a positive reward is given when a target latency is achieved, or a certain number of UEs are successfully served. The state transition probability represents the likelihood of moving from one state to another after taking an action. In the proposed framework, the environment (i.e., the network) evolves dynamically based on user density and the agent’s actions. The quality of an action is evaluated through the reward function. The overall reward is determined based on three criteria: latency-based, UE-based, and action-based rewards. Let the average latency be denoted by \(t_{\mathrm{A}}\), the target latency by \(t_{\mathrm{t}}\), and the penalty factor by \(\mathsf{P}_{\mathrm{f}}\). Then, the latency-based reward \(R_
{\mathrm{t}}\) can be expressed as follows:

\begin{equation} 
\label{deqn_ex5}
R_{\mathrm{t}} =
\begin{cases}
- \mathsf{P}_{\mathrm{f}} (t_{\mathrm{A}} - t_{\mathrm{t}}), & \text{if } t_{\mathrm{A}} > t_{\mathrm{t}} \\
\;\; \mathsf{P}_{\mathrm{f}} (t_{\mathrm{t}} - t_{\mathrm{A}}), & \text{if } t_{\mathrm{A}} \leq t_{\mathrm{t}}
\end{cases}.
\end{equation}

Therefore, a significant penalty is imposed when the latency exceeds the target, while a reward is provided when the latency remains below the target.
The next component is the user-serving reward. Let \(N_{\mathrm{UES}}\) denote the number of served UEs and \(N_{\mathrm{UEB}}\) denote the baseline number of UEs. Then, the UE-based reward function can be found as follows:

\begin{equation} 
\label{deqn_ex6}
R_{\text{UE}} =
\begin{cases}
\mathrm{C}_1 (N_{\text{UES}} - N_{\text{UEB}}), & \text{if } N_{\text{UES}} > N_{\text{UEB}} \\
- \mathrm{C}_2 (N_{\text{UEB}} - N_{\text{UES}}), & \text{if } N_{\text{UES}} < N_{\text{UEB}} \\
0, & \text{if } N_{\text{UES}} = N_{\text{UEB}}
\end{cases}.
\end{equation}

Therefore, a positive reward is assigned when a larger number of UEs are served, while a negative reward is given when the serving capability decreases. The final component is the action-based reward, which considers the addition and removal of DU or RU UAVs. A positive reward is assigned when removing UAVs leads to improved service efficiency, whereas unnecessary additions or removals may incur a penalty. Thus, the overall reward function can be expressed as follows:
\begin{equation} 
\label{deqn_ex7}
R_{\mathrm{a}} =
\begin{cases}
- \mathrm{C}_3, & \text{if a DU and RU UAV is added} \\
\;\; \mathrm{C}_4, & \text{if a DU and RU UAV is removed}
\end{cases}.
\end{equation}

To sum up, the total rewards \(R_{\mathrm{TL}}\) can be obtained as:
\begin{equation} 
\label{deqn_ex8}
R_{\mathrm{TL}}=R_{\mathrm{t}}+R_{\text{UE}}+R_{\mathrm{a}}.
\end{equation}


\begin{table}[!t]
\caption{Simulation Parameters for RL Model Training}
\label{tab:table3}
\centering
\begin{tabular}{ll}
\textbf{Parameters} & \textbf{Values} \\
\hline\hline
\(N_{\mathrm{HAPS}}\), \(N_{\mathrm{LEO}}\), \(N_{\mathrm{UAV}}\), \(N_{\mathrm{UE}}\) & 1, 1, 31, 286 \\ \hline
\(N_{\mathrm{RU}}\), \(N_{\mathrm{DU}}\), \(N_{\mathrm{RL}}\) & 27, 6, 1 \\ \hline
\(\alpha\), \(\gamma\), \(\epsilon\), \(\mathsf{P}_{\mathrm{f}}\) &  0.9, 0.9, 0.9, 1000 \\ \hline
\(\mathrm{C}_1\), \(\mathrm{C}_2\), \(\mathrm{C}_3\), \(\mathrm{C}_4\) & 0.1, 0.5, 5, 5 
  \\ \hline
\(\sigma_{x_i}\), \(\sigma_{y_i}\), \(\mathsf{A}_\mathrm{B}\)  & 800, 800, 5000 \(\mathrm{m^2}\) \\ \hline
\end{tabular}
\end{table}

\section{GNN-enhanced RL Training Performance Analysis}

The simulation and training framework is developed using the Python ecosystem through Anaconda Navigator. All experiments are conducted within Jupyter Notebook to facilitate interactive development and iterative evaluation. The proposed graph-based network model is implemented using NetworkX, while numerical computations and learning processes are supported by NumPy. Visualization of performance metrics, including latency and reward convergence, is performed using Matplotlib. This integrated environment ensures scalability, reproducibility, and efficient prototyping of the GNN-enhanced RL model. All the RL model simulation parameters are presented in Table~III.

\begin{algorithm}[H]
\caption{Q-Learning-based Dynamic NTN}
\label{alg:qlearning_ntn}
\begin{algorithmic}[1]
\STATE Initialize Q-table $Q(s,a)$ arbitrarily
\STATE Set learning rate $\alpha$, discount factor $\gamma$, exploration rate $\epsilon$
\STATE Initialize NTN graph $G$ with initial Data Network, CU GW, \(N_{\mathrm{LEO}}\), \(N_{\mathrm{HAPS}}\), \(N_{\mathrm{UE}}\), \(N_{\mathrm{RU}}\)
, \(N_{\mathrm{DU}}\), \(N_{\mathrm{LN}}\), and \(N_{\mathrm{RL}}\)
\FOR{each episode}
    \STATE Reset environment: $G \leftarrow G_{initial}$
    \STATE Update node positions and compute initial latencies
    \FOR{each step}
        \STATE Observe current state $s = (N_{\mathrm{DU}}, N_{\mathrm{RU}}, N_{\mathrm{UE}})$
        \STATE Select action $a$ using $\epsilon$-greedy policy:
        \IF{rand() $< \epsilon$}
            \STATE Choose random action
        \ELSE
            \STATE $a = \arg\max_a Q(s,a)$
        \ENDIF
        
        \STATE Execute action $a$:
        \IF{$a =$ add\_DU\_UAV and $N_{\mathrm{DU}} < N_{\mathrm{DU}}^{max}$}
            \STATE Add a DU UAV to graph $G$
        \ELSIF{$a =$ add\_RU\_UAV}
            \STATE Select a DU UAV and add RU UAV and $N_{\mathrm{RU}} < N_{\mathrm{RU}}^{max}$
        \ENDIF
        
        \STATE Update node positions and recalculate latencies
        \STATE Observe new state $s'$
        
        \STATE Compute reward $R$ based on latency and user coverage
        
        \STATE Update Q-value:
        \STATE $Q(s,a) \gets Q(s,a) + \alpha \bigl[ R + \gamma \max_{a'} Q(s',a') 
         - Q(s,a) \bigr]$
        
        \STATE $s \leftarrow s'$
    \ENDFOR
    
    \STATE Store metrics
    $(t_{\mathrm{A}}, R_{\mathrm{TL}}, N_{\mathrm{DU}}, N_{\mathrm{RU}}, N_{\mathrm{UE}})$ 
    \STATE Update exploration rate:
    \STATE $\epsilon \leftarrow \max(\epsilon_{min}, \epsilon \times decay)$
    
\ENDFOR

\STATE \textbf{Return} optimal policy $\pi^*(s)$ and learned Q-table
\end{algorithmic}
\end{algorithm}

\begin{figure}[ht]
\centering
\includegraphics[width=3.5in]{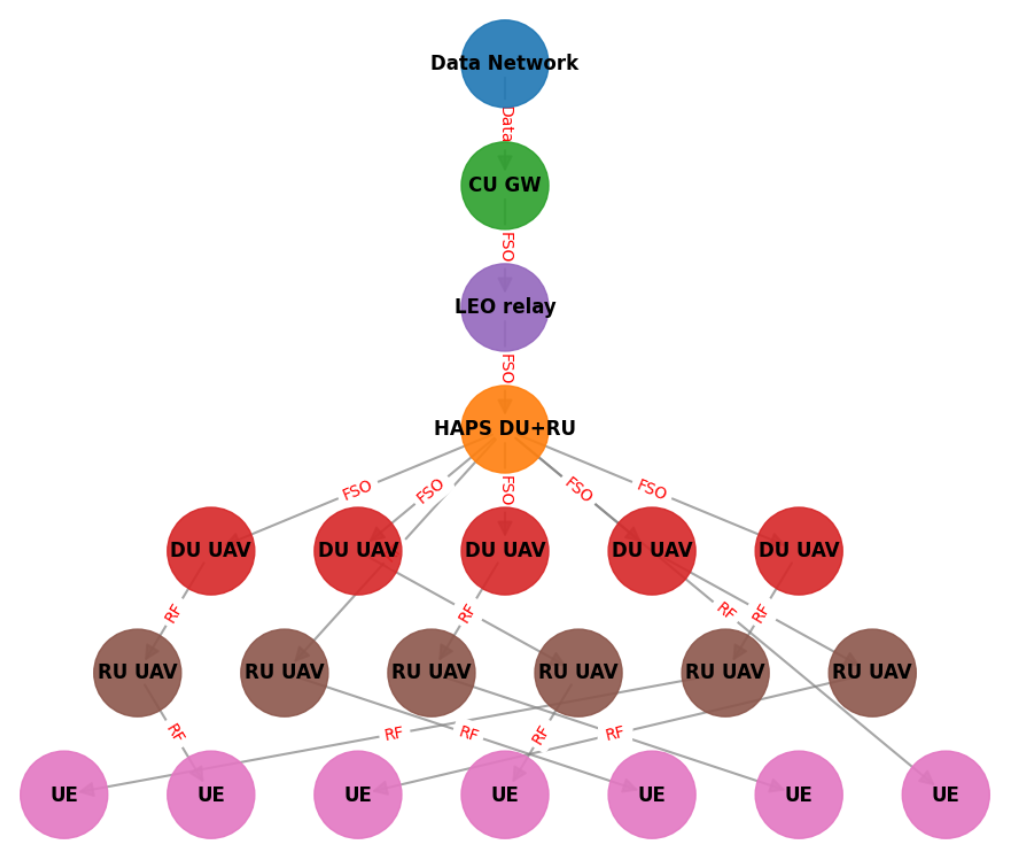}
\caption{GNN hierarchical network topology, smaller unit of the graph.}
\label{fig_2}
\end{figure}

\begin{figure}[!t]
\centering
\includegraphics[width=3.5in]{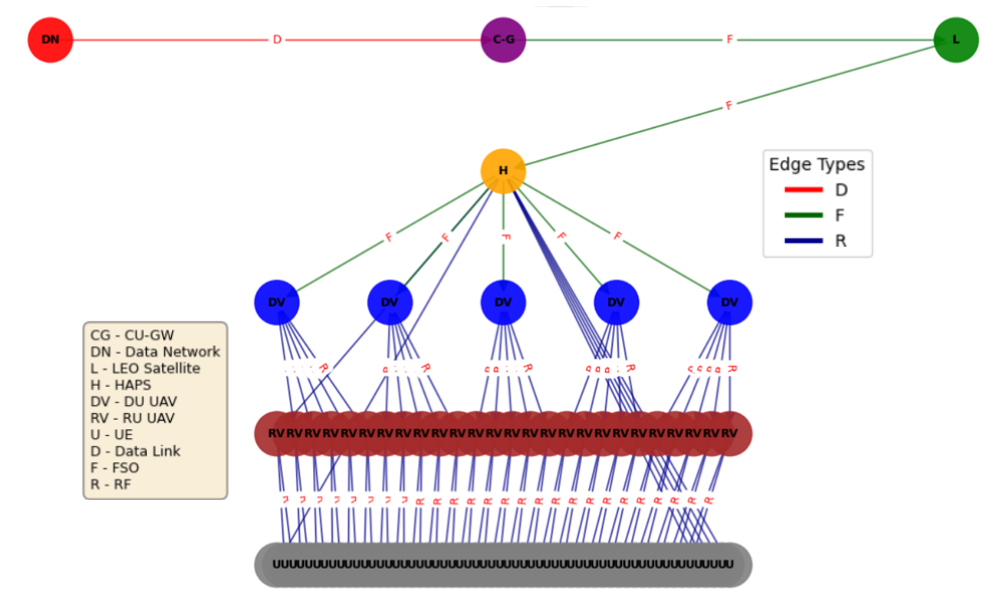}
\caption{GNN hierarchical end-to-end network topology.}
\label{fig_3}
\end{figure}

Fig. 2 depicts a smaller unit of the network, where all fundamental NTN platforms are shown. Fig. 3 presents the complete end-to-end network topology based on the GNN model. The network consists of 286 UEs, 31 UAVs, 27 O-RUs, 6 O-DUs, and one relay unit, in addition to data networks and a CU-GW distributed across different NTN platforms. The RL training model is then executed based on this topology. Fig. \(4\) illustrates the total reward characteristics of the RL-based method as a function of the number of training episodes for different numbers of UEs within the DBS coverage area. From the figure, it can be observed that, as the number of users increases, the total training reward decreases. Moreover, reward fluctuations become more pronounced with a higher number of UEs. As the number of UEs increases, the accuracy of the actions taken to satisfy the target latency declines, leading to reduced rewards. In the scenario with 186 UEs, the reward converges more rapidly and stabilizes within fewer than 100 episodes. Moreover, when the number of users is smaller, fewer DU or RU UAVs are required; consequently, redundant UAVs can be removed, resulting in higher action-based rewards. 

\begin{figure}[!t]
\centering
\includegraphics[width=1\linewidth]{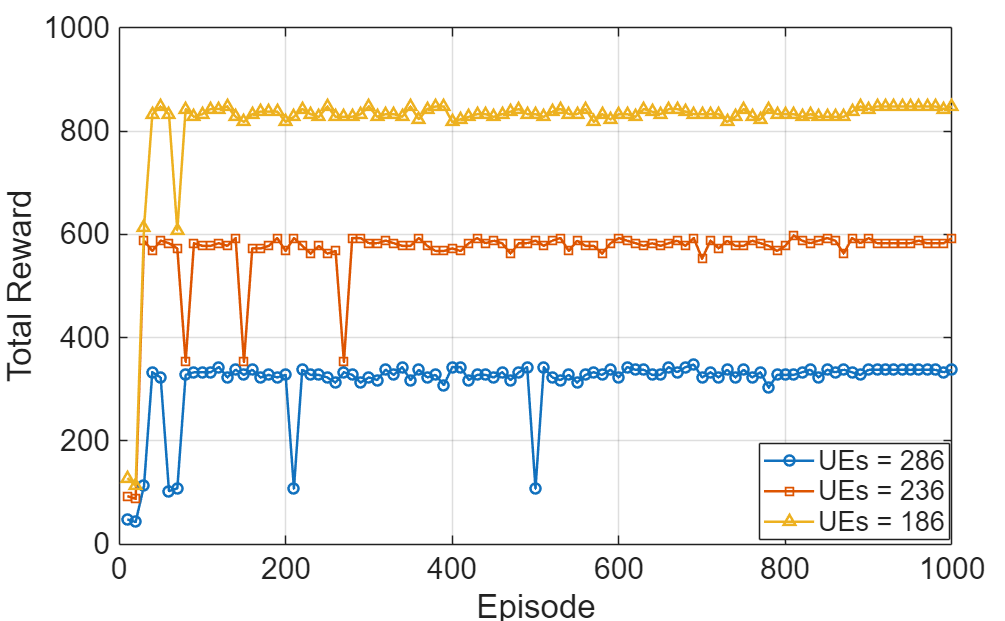}
\caption{Total reward for different numbers of UEs.}
\label{fig_4}
\end{figure}

\begin{figure}[!t]
\centering
\includegraphics[width=1\linewidth]{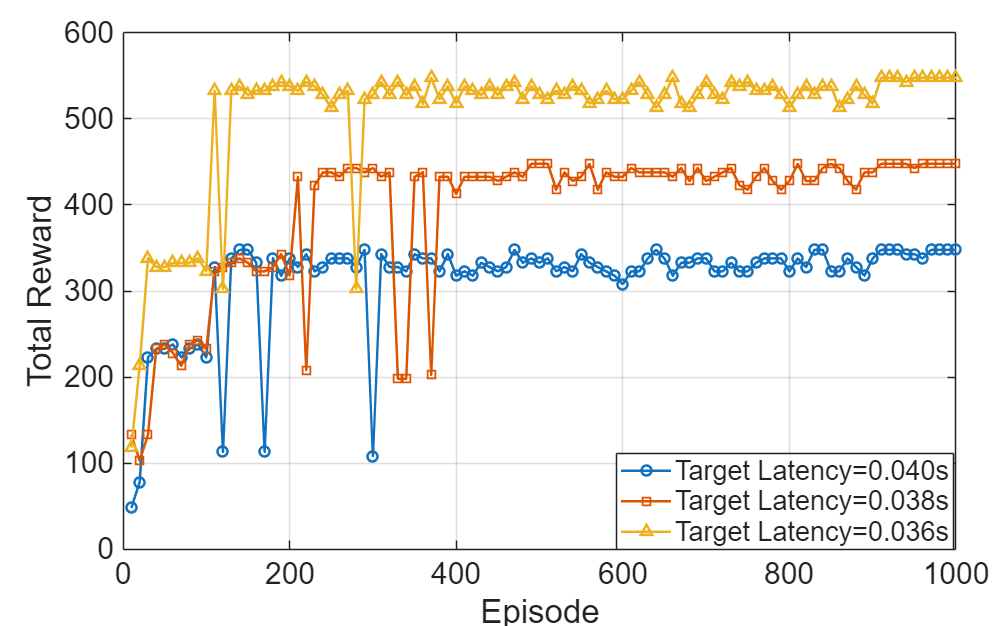}
\caption{Total reward for different latency targets.}
\label{fig_5}
\end{figure}

Fig. \(5\) illustrates the total reward characteristics as a function of the target latency. From the figure, it is evident that, in the early episodes, there are significant reward fluctuations due to increased exploration and exploitation. Moreover, the addition or removal of DU or RU UAVs can cause sudden drops in the total reward during the initial stages. After approximately 400 episodes, the agent becomes well-trained within the environment and exhibits more stable rewards. On the other hand, as the target latency decreases, the total reward increases. This is because a lower target latency creates a greater contrast with the average latency, resulting in higher rewards.

\begin{figure}[!t]
\centering
\includegraphics[width=1\linewidth]{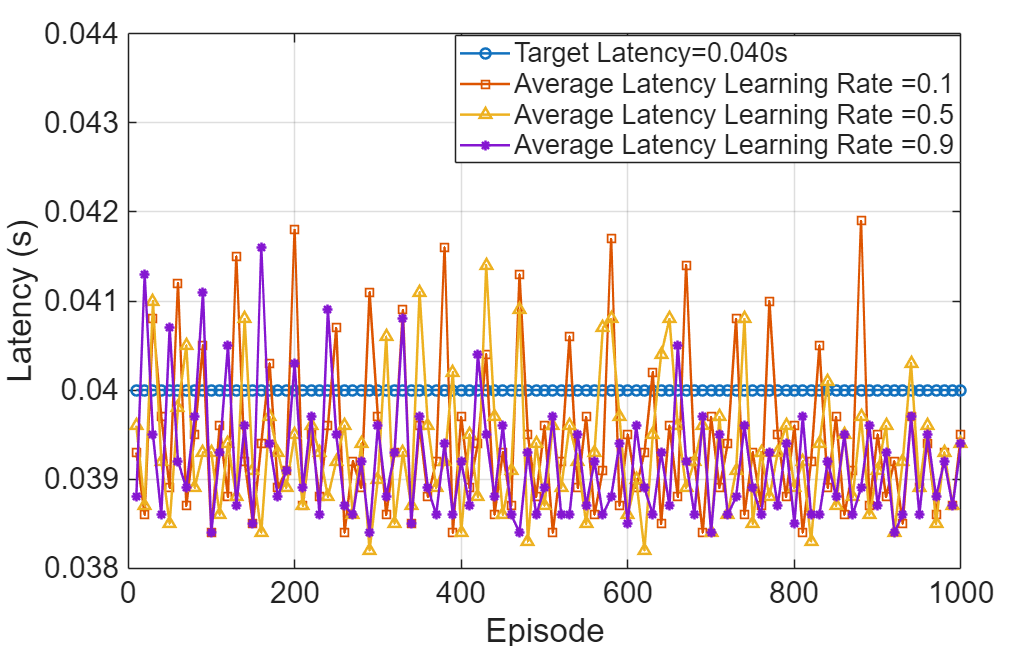}
\caption{Latency with different learning rates.}
\label{fig_6}
\end{figure}

\begin{figure}[!t]
\centering
\includegraphics[width=1\linewidth]{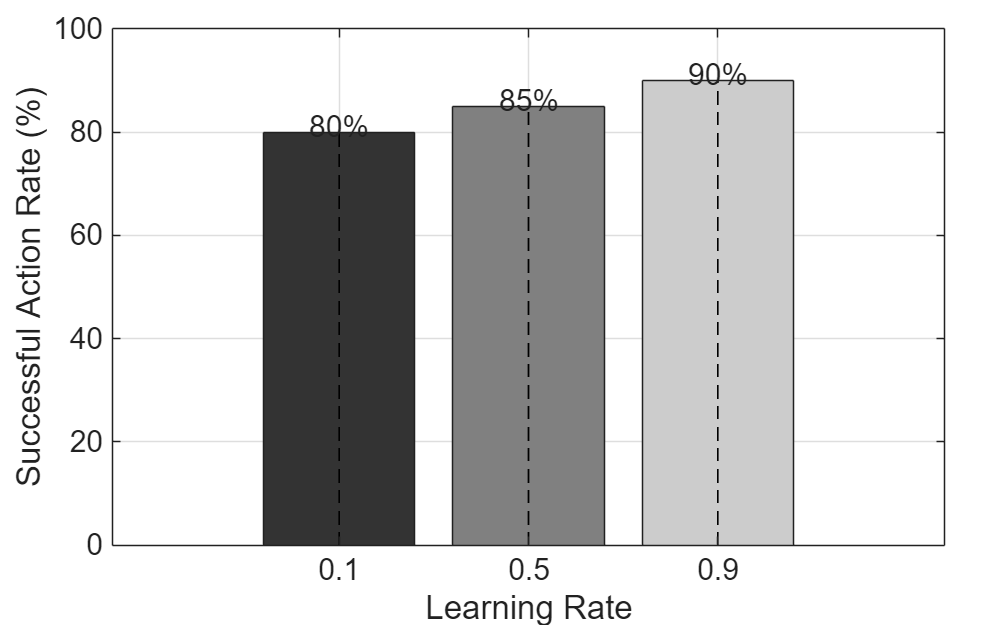}
\caption{Successful action rate for different types of learning criteria.}
\label{fig_7}
\end{figure}

Fig. \(6\) represents the latency distribution as a function of the number of episodes for different learning rates. If the baseline is considered as the target latency, it can be observed that, as the learning rate increases, the latency exceeding the target latency decreases. A higher learning rate enables the agent to learn the environment more quickly; therefore, the deviation from the target latency is reduced. Moreover, as the number of episodes increases, the average latency exceeding the target latency decreases. Based on the results in Fig. 6, Fig. 7 presents a bar chart of the successful action rates. An action is considered unsuccessful if the average latency exceeds the target latency. From the figure, it is evident that the successful action rate increases with the learning rate. This is because a higher learning rate allows the agent to adapt more rapidly to environmental changes and make more accurate decisions, resulting in improved performance.

\section{Conclusion and Future Work}


In this paper, an intelligent ORAN-based NTN emergency network control system has been developed to address the limitations of TNs in next-generation wireless communication scenarios. By integrating a GNN-enhanced RL model, the proposed framework enables efficient and adaptive decision-making for real-time network control, particularly in emergency situations where reliable connectivity is critical. The performance of the proposed RL-based approach has been thoroughly evaluated using key metrics such as total reward, average latency, and successful action rate under different learning rate settings. The results demonstrate that the proposed model achieves a successful action rate exceeding 80\% while maintaining stable latency performance across varying learning configurations and observation conditions. Future work will focus on hyperparameter optimization, improving model generalization, and incorporating additional state representations. Comparative evaluations with baseline methods will be conducted to justify the selected parameters through practical experiments, while real-time UAV-assisted NTN data will be integrated to develop more efficient and scalable algorithms for practical deployment.


\balance
\bibliographystyle{IEEEtran}
\bibliography{reference}



\end{document}